\documentclass[doublecol]{epl2} 
\def\kbar{{\mathchar'26\mkern-9mu k}}
\usepackage{amssymb}
\usepackage{psfrag}
\usepackage{ulem}
\newcommand{\bbm}[1]{\boldsymbol{\mathbf{#1}}}

\title{Universality of the Anderson transition with the quasiperiodic kicked rotor}
\shorttitle{Universality of the Anderson transition} 

\author{G. Lemari\'e\inst{1} \and B. Gr\'emaud\inst{1,2} \and D. Delande\inst{1}}
\shortauthor{G. Lemari\'e \etal}

\institute{                    
  \inst{1} Laboratoire Kastler Brossel, UPMC, ENS, CNRS; 4 Place Jussieu, F-75005 Paris, France\\
  \inst{2} Centre for Quantum Technologies, %
National University of Singapore, 3 Science Drive 2, Singapore 117543, Singapore
}

\pacs{72.15.Rn}{Localization effects (Anderson or weak localization) }
\pacs{03.75.-b}{Matter waves}
\pacs{71.30.+h}{Metal-insulator transitions and other electronic transitions}

\abstract{We report a numerical analysis of the Anderson transition in a quantum-chaotic system, 
the quasiperiodic kicked rotor with three incommensurate frequencies. 
It is shown that this dynamical system exhibits the same critical phenomena as the truly random 3D-Anderson model. 
By taking proper account of systematic corrections to one-parameter scaling, the universality of the 
critical exponent is demonstrated. Our result $\nu = 1.59 \pm 0.01$ is in perfect agreement with the value
found for the Anderson model.}

\begin{document}

\maketitle
\section{Introduction}

It is now widely acknowledged that the classical diffusive behavior of non-interacting electrons in a 
disordered potential can be stopped by non-trivial interference effects \cite{Anderson:PR58}. 
This puzzling phenomenon, Anderson localization, constitutes one strong evidence of the 
very difference between quantum and classical dynamics of complex systems. {A similiar phenomenon is observed 
in the dynamics of the quantum kicked rotor, a paradigmatic system of quantum chaos: the dynamical localization, where quantum mechanical 
interference tend to suppress the classical chaotic diffusive dynamics.} 
The discovery of the parallel between dynamical localization and Anderson localization originated 
from the mapping of the kicked rotor to the quasirandom 1D Anderson model \cite{Fishman:PRL82}. 
In Ref.~\cite{Casati:PRL90} it was demonstrated that the kicked rotor in the dynamical localization 
regime could be modeled by random band matrices; the latter have been reduced to a 
1D non-linear $\sigma$ model \cite{Mirlin:PRL91} similar to those employed in the localization 
theory \cite{Efetov:book97}. In Ref.~\cite{Atland:PRL96} the direct correspondence between 
the kicked rotor and the diffusive supersymmetric non-linear $\sigma$ model was demonstrated. 
In the localized regime, the kicked rotor \textit{exactly} mimics the behavior of disordered electronic conductors.

There has been much experimental efforts to observe Anderson localization in 3D. 
However, due to stray effects like interaction, decoherence or absorption, very few attempts have been successful \cite{Janssen:PhysRep98}. In a slightly different context, Anderson localization of acoustic \cite{Page:NatPhys08} and electromagnetic \cite{Genack:N00,Wiersma:N97,Maret:PRL06,Segev:N07} waves has been experimentally observed. 
The experimental realization of the kicked rotor with laser-cooled atoms interacting 
with a pulsed standing wave allowed for the first experimental observation of Anderson 
localization in 1D with atomic matter waves \cite{Raizen:PRL94}. 
One step further is to observe the well-known Anderson \textit{transition} with this type of system, 
i.e. the disorder induced metal-insulator transition predicted for non-interacting electrons in a 3D disordered potential. 
Different generalizations of the kicked rotor have been theoretically considered 
as analogs of the 3D-Anderson model~\cite{GarciaGarcia:PRE09}. 
Here, we focus on the convenient three-incommensurate-frequencies generalization introduced 
in Ref.~\cite{Shepelyansky:PRL89}. Very recently an experiment based on this system has fully characterized 
the Anderson metal-insulator transition~\cite{AP:Anderson:PRL08}:
a careful analysis of the scaling properties of the dynamics resulted in the first experimental determination
of the localization length critical exponent $\nu$. 
The value found $\nu = 1.4 \pm 0.3$ is compatible with the precedent numerical determination 
of $\nu = 1.57 \pm 0.02$ for the true-random 3D-Anderson model \cite{Slevin:PRL99}.

At this stage, the equivalence between the quasiperiodic kicked rotor \cite{AP:Anderson:PRL08} and 
3D-disordered conductors still has the status of a conjecture (see \cite{Basko:PRL03}). 
A rigorous answer to the question whether this dynamical system exhibits the same 
critical phenomena --i.e. belongs to the same university class--  as the true 3D-Anderson model has not yet been given. 
Can a simple three-frequency dynamical system exactly mimic the critical behavior of 3D disordered electronic conductors? 
In this Letter, we show that the answer is positive. 
This is done by carrying out a very precise numerical study of the critical 
behavior of the quasiperiodic kicked rotor \textit{with the same care} as in the most 
sophisticated investigations of the critical behavior of the true
3D-Anderson model \cite{Slevin:PRL97,Slevin:PRL99,Schreiber:EPJ00}. The fact that both models give 
the same localization length critical exponent $\nu$ within comparable (small) 
uncertainties implies that they belong to the same universality class (orthogonal)~\cite{Evers:RMP08}. 

\section{The quasiperiodic kicked rotor}
The quasiperiodic kicked rotor we consider is a three-incommensurate-frequencies generalization of the kicked rotor:
\begin{equation}\label{eq:KRquasiper}
H_{qp}=\frac{p^{2}}{2}+\mathcal K(t)  \cos \theta \sum_{n}\delta(t-n)\;,
\end{equation}
obtained simply by modulating the amplitude of the standing wave pulses
with a set of two new incommensurate frequencies $\omega_{2}$ and $\omega_{3}$ of modulation:
\begin{equation}\label{eq:Kdet}
\mathcal K (t) = K \left[1+\varepsilon\cos\left(\omega_{2}t+\varphi_{2}\right)\cos\left(\omega_{3}t+\varphi_{3}\right)\right] \; .
\end{equation}
Here we consider the case with an effective Planck constant $\kbar= -i [\theta, p ]$. 
For a standard rotor, $\theta$ is an angle defined modulo $2\pi$ and the wavefunction thus has to be
$2\pi$ periodic. In the atomic realization of the kicked rotor~\cite{AP:Anderson:PRL08}, the Hamiltonian is still given by 
Eq.~(\ref{eq:KRquasiper}), with $\theta$ extended in the $(- \infty , + \infty)$ range.
Using the Bloch theorem, one can restrict to $2\pi$ periodic functions, at the (cheap) price of
including a constant quasi-momentum. 

The dynamics of this quasiperiodic kicked rotor is \textit{identical} to the time-evolution of a 3D-kicked rotor:
\begin{eqnarray}
\lefteqn{ H_{3}=\frac{{p_{1}}^{2}}{2}+\omega_{2}p_{2}+\omega_{3}p_{3} } \nonumber \\
& &+ K\cos\theta_{1}\left[1+\varepsilon\cos \theta_{2} \cos \theta_{3} \right]\sum_{n}\delta(t-n)\;,\label{eq:KR3DquasiperH}
\end{eqnarray}
with an initial condition:
\begin{equation} \label{eq:Psi3}
\Psi_3(\theta_{1},\theta_{2},\theta_{3},t=0) \equiv\Psi_{qp}(\theta_{1},t=0)\delta(\theta_{2}-\varphi_{2})\delta(\theta_{3}-\varphi_{3})\;.\label{eqPsi3}
\end{equation}
where $\Psi_{qp}(\theta,t=0)$ is an arbitrary initial condition for the quasiperiodic kicked rotor. 
Note that dynamical localization takes place in momentum and not in configuration, space. 
The initial state being perfectly localized in $\theta_{2}$ and $\theta_{3}$, it is
entirely delocalized in the conjugate momenta $p_2$ and $p_3$ so that we will study
transport along the $p_1$ direction, which is tantamount to measure the momentum
distribution of the quasiperiodic kicked rotor $\vert \Psi_{qp}(p,t)\vert^2$. The unusual linear dependence of $H_3$ with $p_2$ and $p_3$ does not prevent,
for $\epsilon\neq 0$ the dynamics to be similarly diffusive or localized along the 3 coordinates.
 
The Hamiltonian $H_3$ is invariant under the transformation 
$T: t\rightarrow -t, \bbm{\theta} \rightarrow -\bbm{\theta}, \mathbf{p}\rightarrow \mathbf{p}$, 
i.e. the time reversal in the momentum representation, which is the relevant one for 
dynamical localization (see \cite{Scharf:JPA89,Smilansky:PRL92}). 
In particular, the choice of a non-zero quasi-momentum and non-zero phases $\varphi_2$ or $\varphi_3$ 
does not break time reversal symmetry. The evolution of the states according to 
Hamiltonian (\ref{eq:KRquasiper}) is governed by the operator $U$ (see below, Eq.~(\ref{eq:U})), 
belonging to the Circular Orthogonal Ensemble class~\cite{garcia-garcia:APP112}, with the additional constraint (\ref{eq:Psi3}) at $t=0$; 
this shows that the dynamical properties of the present quasiperiodic kicked rotor also belong to the orthogonal ensemble.

It should also be noted that the 3D aspect comes from the fact that 3 frequencies
are present in our dynamical system: the usual ``momentum frequency''
$\kbar$ which is present in the standard kicked rotor (with $\varepsilon = 0$),
and the two additional time-frequencies $\omega_{2}$ and $\omega_{3}$. By increasing the 
number of incommensurate frequencies, one should be able to tune the effective 
dimensionality of the system. This holds the promise of extending the study of the Anderson transition to higher dimensions.

From a stroboscopic point of view, the quantum dynamics of the 3D-kicked rotor Eq.~(\ref{eq:KR3DquasiperH}) 
is determined by its evolution operator over one period:
\begin{eqnarray}\label{eq:U}
\lefteqn{U=e^{-iK\cos\theta_{1}(1+\varepsilon\cos\theta_{2}\cos\theta_{3})/\kbar} } \nonumber\\
& & \times e^{-i\left(p_{1}^{2}/2+\omega_{2}p_{2}+\omega_{3}p_{3}\right)/\kbar}\;,
\end{eqnarray}
whose eigenstates form a basis set allowing to calculate the temporal
evolution. These Floquet states $\vert\phi\rangle$ are fully characterized
by their quasienergy $\omega$, defined modulo $2\pi$: 
\begin{equation}
U\vert\phi_{\omega}\rangle=e^{-i\omega}\vert\phi_{\omega}\rangle\;.\label{eq:QuasiStates}
\end{equation}
Equivalence with a 3D-Anderson tight-binding
model can be obtained by reformulating Eq.~(\ref{eq:QuasiStates}) for
the Floquet states \cite{Fishman:PRL82}:
\begin{equation}
\epsilon_{\mathbf{m}}\Phi_{\mathbf{m}}+\sum_{\mathbf{r}\neq0}W_{\mathbf{r}}\Phi_{\mathbf{m}-\mathbf{r}}
=-W_{\mathbf{0}}\Phi_{\mathbf{m}}\;,\label{eqAndersonmodelKRquasiper}
\end{equation}
where $\mathbf{m}\equiv(m_{1},m_{2},m_{3})$ and $\mathbf{r}$ label
sites on a 3D lattice, and the $\Phi_{\mathbf{m}}$ are simply related to the Fourier
components of the Floquet state $\vert\phi_{\omega}\rangle.$ 
The on-site energy $\epsilon_{\mathbf{m}}$ reads:
\begin{equation}\label{eq:pseudo-random-disorder-quasiperKR}
\epsilon_{\mathbf{m}}=\tan\left\lbrace \frac{1}{2}\left[\omega-\left(\kbar\frac{{m_{1}}^{2}}{2}
+\omega_{2}m_{2}+\omega_{3}m_{3}\right)\right]\right\rbrace \;,\end{equation}
 and the hopping amplitudes $W_{\mathbf{r}}$ are coefficients of a threefold Fourier expansion
of  $W(\mathbf{\theta}) = \tan\left[K\cos\theta_{1}(1+\varepsilon\cos\theta_{2}\cos\theta_{3})\right/2 \kbar]$.

When {$(\kbar,\omega_{2},\omega_3, 2\pi)$ is an incommensurate quadruplet}, the classical dynamics can become
chaotic (for sufficiently large stochasticity parameter $K \gtrsim 2$) with diffusive spreading 
in \textit{all} $\mathbf{m}$ directions
\cite{Shepelyansky:PRL89}. 
The pseudo-random character
of the potential $\epsilon_{\mathbf{m}}$ then gives the disorder in the model (\ref{eqAndersonmodelKRquasiper}).
{This pseudo-random disorder is not $\delta$-correlated, which implies that the
quasiperiodic kicked rotor is not identical to the Anderson model. It is known that long-range potential correlations may affect
in various ways (including destroy the localization) the Anderson localization~\cite{Tessieri, Carpena}; however,
in our system, these long-range correlations are absent.}

We therefore expect to observe
localization effects as predicted for the standard 3D-Anderson model. 
Localized states would be
observed if the disorder strength is large as compared to the hopping
amplitudes. In the case of model (\ref{eqAndersonmodelKRquasiper}),
while the amplitude of disorder is fixed, the hopping amplitudes increase with $K$. 
Therefore, for large $K$ a diffusive regime should be observed, while for small $K$ 
(though sufficiently large as to prevent classical localization effects) a localized regime is expected. 
This was validated both numerically \cite{Shepelyansky:PRL89} and experimentally \cite{AP:Anderson:PRL08}.

\section{Finite-time scaling}
In the case of the kicked rotor, 
the initial classical chaotic diffusion is stopped only after a certain characteristic time, 
the localization time $\tau_\ell$ which turns out to be roughly proportional to the 
localization ``length'' (characterizing the exponential localization in momentum space) \cite{Shepelyansky:PRL86}: $\tau_\ell \propto \ell$. 
In 3D, the localization time scales as $ \tau_\ell \sim \ell^3$. 
Thus for very large localization length, $\tau_\ell$ may exceed the largest time accessible 
(the maximum duration of a cold-atoms experiment is typically $150$ kicks \cite{AP:Anderson:PRL08} 
whereas numerical investigations can go up to $10^6$ kicks). 
Consequently, it is vain to investigate the Anderson transition only
from static properties of the quasiperiodic kicked rotor, such as the divergence of the localization length at criticality, which could be obtained only for $t \gg \tau_\ell$.
{Note however that some useful information can be extracted from the statistical
properties of the energy levels (the Floquet quasi-energies in our specific case of a kicked system),
which display marked changes at the transitions~\cite{garcia-garcia:APP112,Carpena,Lemarie}.}

Actually, these finite-time effects are similar to finite-size
effects for the study of classical or quantum phase transitions.
We can generalize the usual scaling laws to cover our time-dependent
problem. The single parameter scaling theory, successfully used for
the standard (static) 3D Anderson model \cite{Abrahams:PRL79,MacKinnon:Kramer:PRL81,Pichard:Sarma:JPC81}, 
can be applied to analyze
the dynamics (see \cite{AP:Anderson:PRL08}) and especially to determine the
critical properties of the Anderson transition, i.e. the critical
exponents. 

The dynamics of the quasiperiodic kicked rotor is conveniently studied by considering the time-evolution of the variance of the momentum distribution $\langle p^{2}\rangle$, {the average being taken over several initial conditions, which corresponds to an average over disorder}.
We can make the following scaling hypothesis for this quantity \cite{Stauffer:94}:
\begin{equation}
\langle p^{2}\rangle=t^{k_{1}}F\left[\left(K-K_{c}\right)t^{k_{2}}\right]\;,\label{eq:scalinglaw}
\end{equation}
 with $F$ a function characteristic of the transition (to be determined)
 and $k_{1}$ and $k_{2}$ two
exponents which can be constrained as explained in the following. 
$K_c$ is the critical value of the stochasticity parameter.  
We must recover as $t \rightarrow \infty$ either a diffusive behavior $\langle p^{2}\rangle \sim D t $ when $K > K_c $ 
or a localized dynamics $\langle p^{2}\rangle \sim {\ell}^2 $ when $K < K_c$. 
In the vicinity of the transition, the diffusion constant $D \sim (K-K_c)^s $ vanishes with the critical exponent $s$ and 
the localization length $\ell \sim (K_c - K)^{-\nu}$ diverges with the critical exponent $\nu$. 
{The} Wegner's law \cite{Wegner:ZFP76} $s=\nu$ in three dimensions, 
{then} leads to $k_1=2/3$ and $k_2 = 1/3\nu$~\cite{Ohtsuki:JPSJ97}.

Therefore, to analyze the scaling of the dynamics, the behavior of the quantity 
$\Lambda =\langle (p/\kbar)^{2} \rangle t^{-2/3}$ should be investigated as a function of time and 
for various stochasticity parameters. 
Such a study was undertaken in \cite{AP:Anderson:PRL08} and allowed for a successful 
experimental characterization of the Anderson transition. 
In order to tackle the problem of universality of the critical behavior, we have to go one step further and 
study whether the critical exponent $\nu$ changes when parameters such as $\kbar, \omega_2, \omega_3$ are modified.

The following discussion is rather intricate but cannot be avoided in a rigorous study. 
Indeed, to reliably distinguish the different universality classes of the Anderson transition 
requires a very precise determination of the critical exponent; for instance, the value 
$\nu = 1.43 \pm 0.04$ for the unitary symmetry class is very close to the one for the 
orthogonal symmetry class \cite{Slevin:PRL97}. For a very accurate determination of $\nu,$ 
possible systematic deviations to one-parameter scaling must be taken into account.

\section{Systematic corrections to scaling}

\begin{figure}
\begin{centering}
\psfrag{x}{$K$ } \psfrag{y}{$\ln\Lambda$ } \includegraphics[width=6cm]{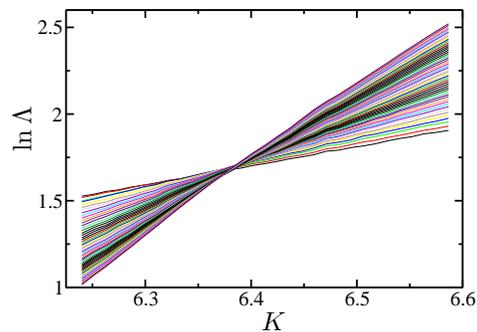} 
\par\end{centering}
\caption{\label{fig:simulambdavsK} (Color online) Dynamics of the quasiperiodic kicked rotor
in the vicinity of the critical regime. 
The rescaled quantity $\ln\Lambda(K,t)$
is plotted as a function of $K$ for various values of time $t$
ranging from $t=30$ to $t=40000$. The
crossing of the different curves at a common point $(K_{c}\simeq6.4,\ln\Lambda_{c}\simeq1.6)$ indicates the 
occurrence of the metal-insulator transition.
The parameters are that of the set $\mathcal A$ (see Table~\ref{Tab:set-of-parameters}).}
\end{figure}

Let us consider the scaling function $\mathcal{F}\equiv\ln (F{/\kbar^2)}$ in the vicinity of the critical point: 
\begin{equation}\label{eq:scalinglnlambda}
\ln\Lambda  =  \mathcal{F}\left[\left(K-K_{c}\right)t^{1/3\nu}\right]\;.
\end{equation}
One simple feature of the scaling hypothesis Eq.~(\ref{eq:scalinglnlambda}) is that when
$\ln\Lambda$ is plotted against $K$, the curves for different times
$t$ should intersect at {the} common point $(K_{c},\ln\Lambda_{c}{=\mathcal{F}(0)})$; this crossing,
which indicates the occurrence of the metal-insulator transition,
is clearly visible in Fig.~\ref{fig:simulambdavsK}.

In practice, the data do not exactly follow Eq.~(\ref{eq:scalinglnlambda}).
There are small systematic deviations to scaling. Here, we consider
several ways in which such corrections can arise (see below for a detailed discussion of each correction): 
(i) the presence of an irrelevant scaling variable, (ii) non-linear dependence of the scaling variables in the stochasticity parameter $K$ and (iii) resonances due to the periods being well approximated by a ratio of small integers.
(iii) is specific to our three-frequency dynamical system, but (i)
and (ii) were shown to play an important role in the standard Anderson
model \cite{Slevin:PRL99}. Note that the most important
correction to scaling is a time dependence of $\ln\Lambda$ at $K=K_{c}$
either due to (i) or (iii).

The well-known deviations (i) and (ii) can be taken into account by adding extra terms
to (\ref{eq:scalinglnlambda})~\cite{Slevin:PRL99}. (i) The scaling function $\mathcal{F}$ depends not only on the relevant
scaling variable $\chi_r$ (i.e. a function of $K-K_c$), but also on an irrelevant scaling
variable $\psi$:
\begin{equation}\label{eq:scalingcorr}
\ln\Lambda=\mathcal{F}\left(\chi_r t^{1/3\nu},\psi t^{-y}\right)\;.
\end{equation}
Since $\psi$ is an irrelevant scaling parameter, its effects should vanish as $t$ goes to infinity, thus $y$ must be positive. 
(ii) Non-linearity in the relevant scaling variable $\chi_r$ can be described by an expansion in terms 
$K-K_{c}$ up to order $m_{R}$.

To define a fitting function $\mathcal{F}_{f}$, we can then make a Taylor expansion of the scaling function $\mathcal F$ up to
order $n_{R}$ in $\chi_r t^{1/3\nu}$ and $n_{I}$ in $\psi t^{-y}$:
\begin{eqnarray}\label{eqmodelSlevin}
\mathcal{F}_{f}(K,t) =  \sum_{m=0}^{n_{R}}\sum_{n=0}^{n_{I}}\chi_r^{m}t^{m/3\nu}\psi^{n}t^{-ny}\mathcal{F}_{m,n} \;.
\end{eqnarray}

\begin{figure}
\begin{centering}
\psfrag{x}{$t$ } \psfrag{y}{$\ln\Lambda$ } \includegraphics[width=6cm]{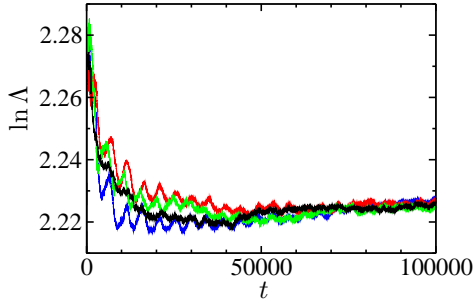} 
\par\end{centering}
\caption{\label{fig:oscillambda} (Color online) 
Small oscillating corrections to the scaling
behavior in the critical regime. The parameters are the following:
$\kbar=2.89$, $\omega_{2}=2\pi\sqrt{5}$, $\omega_{3}=2\pi\sqrt{13}$,
$K=7.8$ and $\varepsilon=0.3$. Color curves correspond to various
choices of the phases $\varphi_{2}$ and $\varphi_{3}$ [see Eq.~(\ref{eq:KRquasiper})]
whereas the black curve results from a statistical average over different
phases. The amplitudes of the quasi-resonant oscillations decrease
as time goes on. Averaging over the phases kills the rapidly oscillating
structures, while keeping all of the other dynamical properties unchanged.}
\end{figure}

We now discuss the qualitative nature of
the correction due to the presence
of an irrelevant scaling variable (i) as compared to the effect of resonances (iii). In case (i), $\ln\Lambda_{c}=\ln\Lambda(K=K_c)$ will shift in a 
monotonous way as time increases and converge to its thermodynamic limit 
$\ln\Lambda_{c}(t=\infty)$ [$\ln\Lambda_{c}(t)=\mathcal{F}_{0,0}+t^{-y}\mathcal{F}_{0,1}$ 
in the linear regime ($m_{R}=n_{R}=n_{I}=1$)]. In the case of our three-frequency
dynamical system, the data do not always fit such a monotonous evolution
model for $\ln\Lambda_{c}(t)$. Indeed, for a generic choice of incommensurate
periods, the data are found to oscillate around their transient
anomalous diffusive dynamics, see Fig.~\ref{fig:oscillambda}. 

We infer that such an oscillating correction to scaling arises in a resonant way: when the frequencies can
be approximately related by a simple linear combination (i.e. involving small integers), a resonance occurs. 
This is a rather common phenomenon in multi-frequency dynamical systems.
The phase and the amplitude of the oscillations in Fig.~\ref{fig:oscillambda} depend on the choice of the phases $\varphi_{2}$
and $\varphi_{3}$ of the time-modulation [see Eq.~(\ref{eq:Kdet})], i.e. on the
initial state in Eq.~(\ref{eq:Psi3}).
From the point of view of the Anderson-like model Eq.~(\ref{eqAndersonmodelKRquasiper}), resonances
can be interpreted as correlations in the disordered potential. 
Hence, to perform the standard scaling analysis
devised for the Anderson model with uncorrelated disorder \cite{Slevin:PRL99},
we shall retain data only for sufficiently long times {(say $t\geq1000$)} and 
average over different initial conditions, i.e.
different quasi-momenta and phases $\varphi_{2}$ and $\varphi_{3}$.

We computed $\ln\Lambda$ for times up to $t=10^{6}$ kicks
with an accuracy of $0.15\%$. To achieve this
accuracy more than $1000$ initial conditions are required. 
To analyze data over the
full range of times $t \in \left[10^{3},10^{6}\right]$, we fit the model
Eq.~(\ref{eqmodelSlevin}) to the data. Note that the inclusion of
the corrections (i) and (ii) in Eq.~(\ref{eqmodelSlevin}) leads to a
rapid increase in the number of fitting parameters. That is why high
quality data with a wide range of variation of $t^{1/3}$ are needed
if meaningful fits are to be obtained. 

\begin{figure}
\begin{center}
\begin{picture}(5,3)
\psfrag{x-inset}{\begin{small}$t$ \end{small} } \psfrag{y-inset}{\begin{small}$\ln\Lambda-\mathcal F_f$\end{small} }
\put(40,100){\includegraphics[width=3cm]{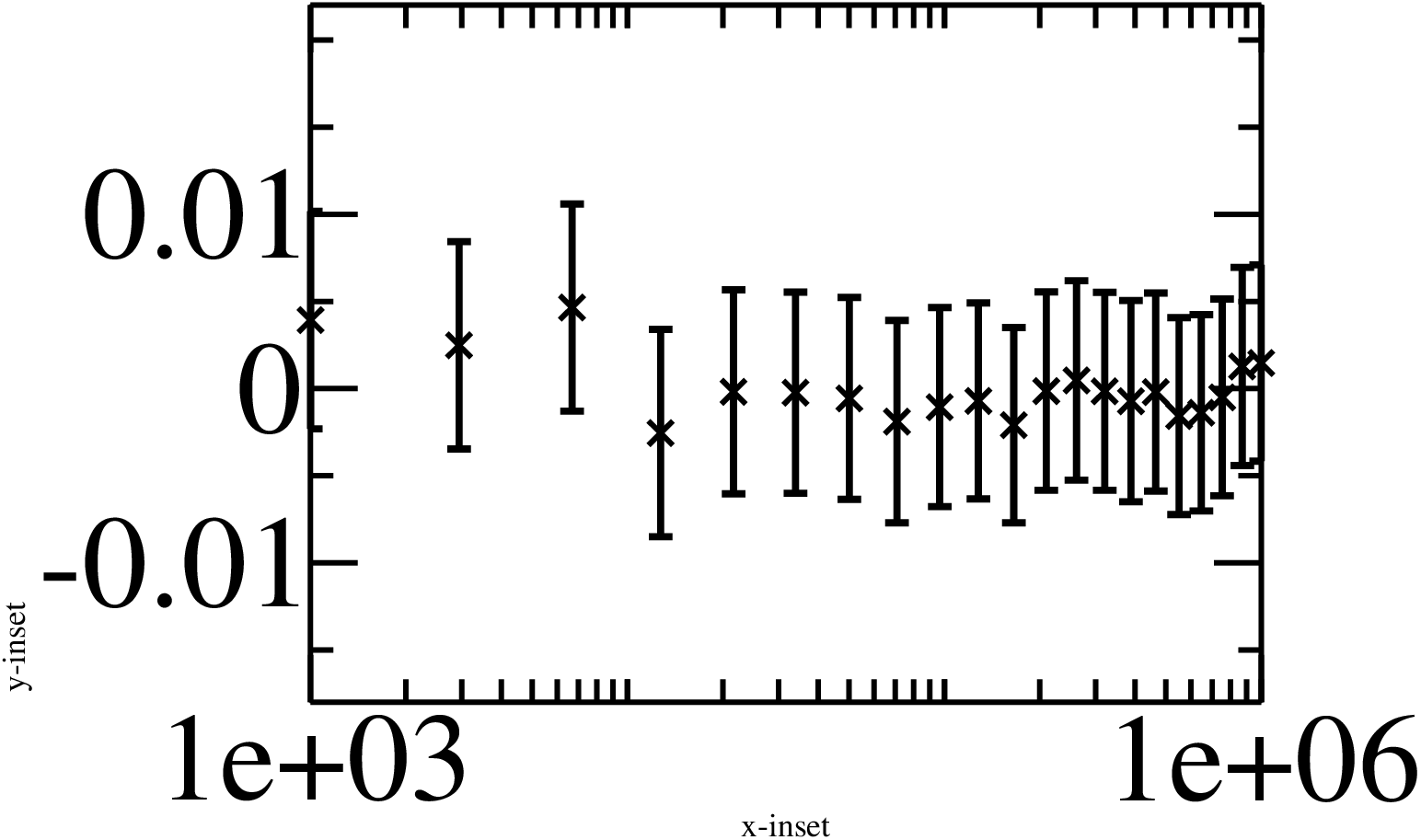}}
\end{picture}
\psfrag{x}{$t$ } \psfrag{y}{$\ln\Lambda$ }
\includegraphics[width=8cm]{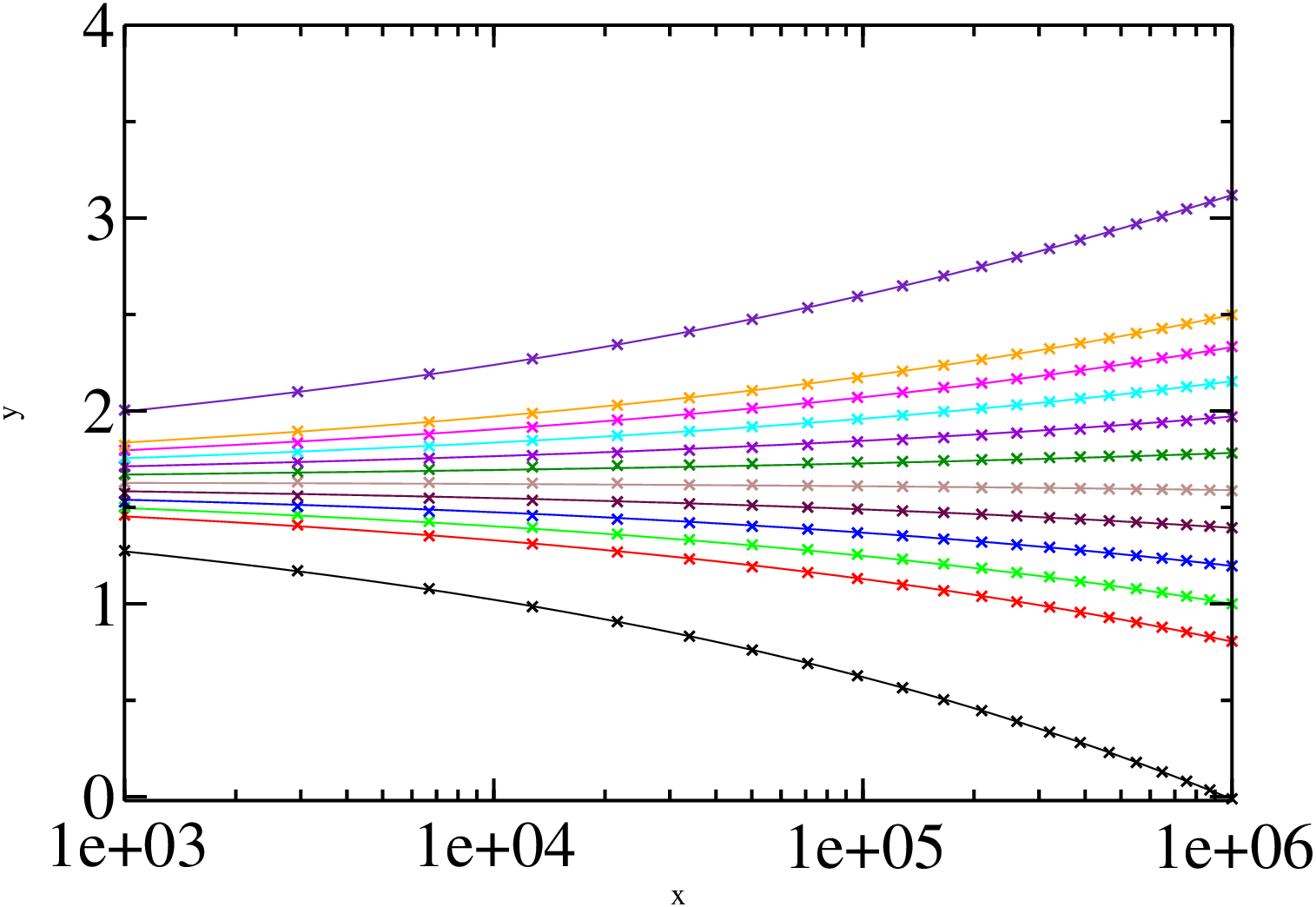}
\caption{\label{fig:scalinglambdavst} (Color online) $\ln\Lambda$ as a function of time
$t$. The \textbf{curves} are fits of the data according to the model $\mathcal{F}_{f}$
{[}see Eq.~(\ref{eqmodelSlevin})] with $n_{R}=3$, $m_{R}=2$, $n_{I}=1$. 
The parameters are that of the set $\mathcal D$ (see Table~\ref{Tab:set-of-parameters}). 
The inset shows the deviations of the data (corresponding to $K=7.9$) to the most likely fit, showing no statistically significant 
deviation.}
\end{center}
\end{figure}

\begin{figure}
\begin{centering}
\psfrag{x}{$\ln\left(\xi/t^{1/3}\right)$ } \psfrag{y}{$\ln\Lambda_{s}$
} \includegraphics[width=7cm]{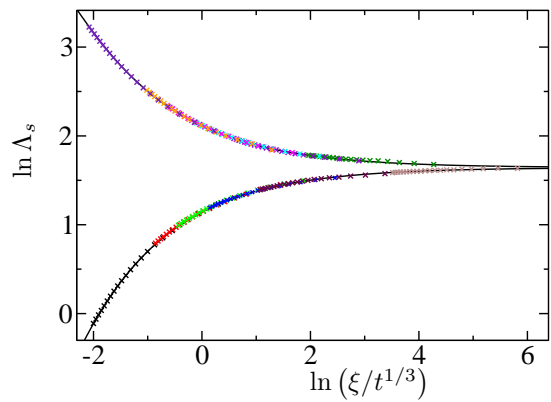} 
\par\end{centering}
\caption{\label{fig:scalinglambdavsxsit13} (Color online) $\ln\Lambda_{s}$, the data in Fig.~\ref{fig:scalinglambdavst} after subtraction of corrections due to
the irrelevant scaling variable, plotted versus $\ln \left(\xi/t^{1/3}\right)$ where $\xi = \vert \chi_r \vert^{- \nu}$ and the scaling function deduced from the model $\mathcal{F}_{f}$, Eq.~(\ref{eqmodelSlevin})
(black curve). The parameters are that of the set $\mathcal D$ (see Table~\ref{Tab:set-of-parameters}). 
The best fit estimates of the critical stochasticity and the critical exponent are: 
$K_c = 8.09\pm 0.01,\ \ln \Lambda_c = 1.64\pm 0.03$ and $\nu = 1.59 \pm 0.01$.}
\end{figure}

The most likely fit is determined by minimizing the $\chi^{2}$ 
statistics measuring the deviation, due to the numerical uncertainties, between the model and data. Some typical numerical data and the associated
fit are displayed in Fig.~\ref{fig:scalinglambdavst}. To exhibit scaling, we subtract the corrections
due to the irrelevant scaling variable \cite{Slevin:PRL99} obtaining the corrected 
quantity $ \ln \Lambda_s$. As seen in Fig.~\ref{fig:scalinglambdavsxsit13}, all data collapse almost
perfectly on the scaling function deduced from the model $\mathcal{F}_{f}$.

No significant deviation of the scaling function from the fit is observed.
The goodness of fit $Q$ has been determined using the
$\chi^{2}$ distribution with $N_d-N_{p}$ degrees of freedom where
$N_d$ is the total number of data we used to fit the model  and $N_p$ 
is the total number of fitting parameters (see Table~\ref{Tab:set-of-parameters}). The confidence intervals {(one standard deviation)} for the fitted parameters were estimated using the bootstrap method which yields Monte Carlo estimates
of the errors in the fitted parameters. 

\section{Universality}

\begin{table}[t]
\begin{centering}
  \begin{footnotesize}  
\begin{tabular}{ c || c| c| c| c| c }

 & $\kbar$ & $\omega_2$ & $\omega_3$ & $K$        & $\epsilon$ \\ \hline\hline
$\mathcal A$ & $2.85$ & $2 \pi \sqrt{5}$ & $2 \pi \sqrt{13}$ & $6.24 \rightarrow 6.58$ &  $0.413 \rightarrow 0.462$
\\ \hline
$\mathcal B$ & $2.85$ & $2 \pi \sqrt{7}$ & $2 \pi \sqrt{17}$ & $5.49 \rightarrow 5.57$ &  $0.499 \rightarrow 0.514$\\ \hline
$\mathcal C$ & $2.2516$ & $1/\eta$ & $1/\eta^2$ &  $ 4.98\rightarrow 5.05$ &  $0.423 \rightarrow 0.436$
\\ \hline
$\mathcal D$ & $3.5399$ & $\kbar/\eta$ & $\kbar/\eta^2$ &  $ 7.9\rightarrow 8.3$ &  $0.425 \rightarrow 0.485$
\end{tabular}
\end{footnotesize}
\par\end{centering}
\caption{\label{Tab:set-of-parameters} 
The four sets of parameters considered: $\kbar$, $\omega_2$ and $\omega_3$ control the 
microscopic details of the disorder, while $\epsilon$ drives the anisotropy of the hopping amplitudes. 
In $\mathcal C$, $\omega_2/\omega_3=\eta$ where $\eta$ is the silver number (see below), 
and the continuous fraction of $\kbar/\pi$ is constituted of small integers. 
$\mathcal D$ is such that $\kbar = \alpha$, $\omega_2 = \alpha / \eta$ and $ \omega_3 = \alpha / \eta^2$ where $\eta=1.324717...$ 
the real root of the equation $\eta^3 - \eta - 1 = 0$, and $\alpha$ 
is such that the continuous fractions of $\kbar/\pi = \alpha/\pi$, $\omega_2/\pi = \alpha/\eta\pi$ and 
$\omega_3/\pi = \alpha/\eta^2 \pi$ are constituted of small ($<9$) integers \cite{Kim:PRA86}.}
\end{table}

A key property of the Anderson transition is that it is
a continuous (i.e. second order) phase transition whose 
critical behavior can be described in a framework of universality classes \cite{Efetov:book97}. This means that the critical
exponents should not be sensitive to the microscopic details of the disordered potential but should
depend only on the underlying symmetries (e.g. time reversal symmetry). We present here new material that allows us to 
numerically prove that this is indeed the case for the quasiperiodic kicked rotor,
thus strengthening the fact that this system is in the same universality class than
the truly random Anderson transition.

We have carried on a \textit{detailed} study of four cases characterized by different set of parameters, 
see Table~\ref{Tab:set-of-parameters}): 
$\mathcal A$ and $\mathcal B$ are both optimal set of parameters for experimental studies (see \cite{AP:Anderson:PRL08}), 
while $\mathcal C$ and $\mathcal D$ are rather for theoretical/numerical considerations. 
$\mathcal C$ is a first step towards an ideal choice of parameters: $\omega_2/\omega_3=\eta$ 
where $\eta$ is the silver number (see below), and the continuous fraction of $\kbar/\pi$ 
is constituted of small integers (to prevent the system to be close to a resonance). 
$\mathcal D$ should be a ``best choice'' of parameters if we seek the least correlations in the 
disorder Eq.~(\ref{eq:pseudo-random-disorder-quasiperKR}). It is such that $\kbar$, $\omega_2$, $\omega_3$ and $\pi$ are 
a ``most incommensurate'' \textit{quadruplet} of numbers. 
As suggested in \cite{GarciaGarcia:PRE09}, 
we set $\kbar = \alpha$, $\omega_2 = \alpha / \eta$ and $\omega_3 = \alpha / \eta^2$ where $\eta=1.324717...$ 
(the silver number which generalizes the golden number for triplet instead of pair of incommensurate numbers \cite{Kim:PRA86}) 
is the real root of the equation $\eta^3 - \eta - 1 = 0$, and $\alpha=3.5399... $ 
is such that the continuous fractions of $\kbar/\pi = \alpha/\pi$, $ \omega_2/\pi = \alpha/\eta\pi$ and $ \omega_3/\pi = \alpha/\eta^2 \pi$ 
are constituted of small ($<9$) integers.

\begin{table}
\begin{centering}
  \begin{footnotesize}  
  \begin{tabular}{ c || c| c| c| c | c |c |c  }
  & $n_R$ & $n_I$ & $m_R$      & $N_d$ & $N_p$ &$\chi_r^2$ & $Q$ \\ \hline \hline
$\mathcal A$ & $3$ & $1$ & $2$  & $800$ &  $12$ & $236$ & $1$ 
\\ \hline
$\mathcal B$ & $2$ & $1$ & $1$   & $600$ &  $9$ & $278$ & $1$ 
\\ \hline
$\mathcal C$ & $2$ & $1$ & $1$   & $1000$ &  $9$ & $934$ & $0.9$ 
\\ \hline
$\mathcal D$ & $3$ & $1$ & $2$    & $1212$ &  $12$ & $917$ & $0.999$ 
\end{tabular}
\end{footnotesize}
\par\end{centering}
\caption{\label{Tab:fit} Parameters used for the four sets of data:  
type of fit [see Eq.~(\ref{eqmodelSlevin})], number of data points $N_d$, 
number of parameters $N_p$, value of $\chi_r^2$ for the best fit and goodness of fit $Q$. 
Time ranges from $t=10^3$ to $t=10^6$.}
\end{table}

\begin{table}[!ht]
\begin{centering}
  \begin{footnotesize}  
  \begin{tabular}{ c || c| c| c| c}
  & $K_c$ & $\ln \Lambda_c$ & $\mathbf \nu$ & $y$ \\ \hline \hline
$\mathcal A$ & $6.36 \pm 0.02$ & $1.60 \pm 0.04$ & $\mathbf{1.58 \pm 0.01}$ & $0.71 \pm 0.28$
\\ \hline
$\mathcal B$ & $5.53 \pm 0.03$ & $1.08 \pm 0.09$ & $\mathbf{1.60 \pm 0.03}$ & $0.33 \pm 0.30$
\\ \hline
$\mathcal C$ & $5.00 \pm 0.03$ & $1.19 \pm 0.15$ & $\mathbf{1.60 \pm 0.02}$ & $0.23 \pm 0.29$
\\ \hline
$\mathcal D$ & $8.09 \pm 0.01$ & $1.64 \pm 0.03$ & $\mathbf{1.59 \pm 0.01}$ & $0.43 \pm 0.23$
\end{tabular}
\end{footnotesize}
\par\end{centering}
\caption{\label{Tab:critical-parameters} 
Best fit estimates of the critical parameters $K_c$ and $\ln \Lambda_c$, the critical exponent $\nu$ together with
their uncertainty (one standard deviation). 
$\nu$ is expected to be universal whereas $\ln \Lambda_c$ and $K_c$ do depend on 
anisotropy \cite{Evangelou:PRB94} and $\kbar$, $\omega_2$ and $\omega_3$. 
Irrelevant parameters are sensitive to microscopic details, therefore $y$ is strictly positive and not universal.}
\end{table}

The details of the simulations and the types of fit used to analyze those sets are listed in Table~\ref{Tab:fit}. 
The estimated critical parameters and their confidence intervals are given in Table~\ref{Tab:critical-parameters}. 
Some \textit{typical} data and scaling function are drawn in Figs. \ref{fig:scalinglambdavst} and \ref{fig:scalinglambdavsxsit13}.

The most important conclusion to be drawn from Table~\ref{Tab:critical-parameters} is that the estimates 
of the exponent $\nu$ for the four different sets are in almost perfect agreement with each other 
and with the estimate of $\nu$ based on numerical studies of the truly random Anderson model $\nu = 1.57 \pm 0.02$ 
of the orthogonal symmetry class \cite{Slevin:PRL99}. Note also that in the case of the quasiperiodic kicked rotor, 
the critical stochasticity $K_c$ and $\ln \Lambda_c$ depend on: (i) the anisotropy governed by the parameter $\epsilon$ and 
(ii) $\kbar$, $\omega_2$ and $\omega_3$. The dependence (i) of the critical disorder and critical $\ln \Lambda$ on 
anisotropy is a typical feature of the Anderson transition in anisotropic solids \cite{Schreiber:EPJ00}. 
The dependence (ii) follows from the relation between the initial ``classical'' diffusion constant and 
the parameters $\kbar$, $\omega_2$ and $\omega_3$ \cite{Shepelyansky:PRL86}.

The Anderson transition with the quasiperiodic kicked rotor is a robust feature: indeed, 
the ``naive'' choices of parameters, set $\mathcal A$ and $\mathcal B$, lead to very clean 
critical behaviors, as clean as for the sophisticated choices of parameters, 
sets $\mathcal C$ and $\mathcal D$. Our experience is that for certain mutually incommensurate triplets 
($\kbar$, $\omega_2$, $\omega_3$) systematic deviations to scaling (such as resonances) 
can occur for intermediate times, but eventually vanish.

\section{Conclusion}
A numerical analysis of the critical behavior of the quasiperiodic kicked rotor
has shown that this quantum-chaotic system exhibits the same critical phenomena as the truly random Anderson model, 
i.e. both systems belong to the same (orthogonal) universality class~\cite{Evers:RMP08}. 
By taking proper account of corrections to the scaling property around criticality, 
the universality of the critical exponent $\nu$ for the quasiperiodic kicked rotor was demonstrated. 
The critical exponent $\nu$ was determined with an accuracy better or comparable to the one achieved 
in previous numerical studies of the 3D-Anderson model \cite{Slevin:PRL97,Slevin:PRL99,Schreiber:EPJ00}. 
Our result $\nu = 1.59 \pm 0.01$ is in perfect agreement with the value found for the orthogonal symmetry class \cite{Slevin:PRL99}. 
It is clearly distinct from the value found for the unitary class $\nu = 1.43\pm 0.04$ \cite{Slevin:PRL97}, 
and from the predictions $\nu = 1$ of the self-consistent theory of localization \cite{Wolfle:PRL82} 
and $\nu = 1.5$ of a recent \textit{ad hoc} refinement of the self-consistent theory \cite{GarciaGarcia:PRL08}.

\acknowledgments
We thank J.C. Garreau, P. Szriftgiser, J. Chab\'e, H. Lignier and F. Farago for many interesting and fruitful discussions.

\end{document}